\documentclass[twocolumn]{aastex63}

\usepackage{apjfonts}
\usepackage{color}
\usepackage{soul}
\usepackage{natbib}
\usepackage{hyperref}
\usepackage{amsmath}
\usepackage{xspace}
\usepackage{graphicx}
\usepackage{enumitem}
\usepackage{amssymb}
\usepackage{xifthen}
\usepackage[normalem]{ulem}
\usepackage{aas_macros}
\usepackage{amsfonts}
\usepackage{amsmath}
\usepackage{multirow}
\usepackage{dsfont}
\usepackage{hyperref}

\newcommand{\code}[1]{\texttt{#1}\xspace}

\hypersetup{    
  colorlinks      = {true},
  linkcolor       = {blue},
  citecolor       = {blue},
  urlcolor        = {blue},
}

\shorttitle{Cetus~II abundances}
\shortauthors{Webber et al.}

\reportnum{DES-2023-0757}
\reportnum{FERMILAB-PUB-23-373-PPD}

\begin{document}
\title{Chemical Analysis of the Brightest Star of the Cetus~II Ultra-Faint Dwarf Galaxy Candidate\footnote{This paper includes data gathered with the 6.5m Magellan Telescopes located at Las Campanas Observatory, Chile.}}

\author[0000-0002-9762-4308]{K.~B.~Webber}
\affil{Mitchell Institute for Fundamental Physics and Astronomy and Department of Physics and Astronomy, Texas A\&M University, College Station, TX 77843-4242, USA}

\author[0000-0001-6154-8983]{T.~T.~Hansen}
\affil{Department of Astronomy, Stockholm University, AlbaNova
University Center, SE-106 91 Stockholm, Sweden}

\author{J.~L.~Marshall}
\affil{Mitchell Institute for Fundamental Physics and Astronomy and Department of Physics and Astronomy, Texas A\&M University, College Station, TX 77843-4242, USA}

\author{J.~D.~Simon}
\affil{Observatories of the Carnegie Institution for Science, 813 Santa Barbara St., Pasadena, CA 91101, USA}
\author{A. ~B. ~Pace}
\affil{McWilliams Center for Cosmology, Carnegie Mellon University, 5000 Forbes Ave, Pittsburgh, PA 15213, USA}
\author{B. ~Mutlu-Pakdil}
\affil{Department of Physics and Astronomy, Dartmouth College, Hanover, NH 03755, USA}

\author{A. ~Drlica-Wagner}
\affil{Fermi National Accelerator Laboratory, P. O. Box 500, Batavia, IL 60510, USA}
\affil{Kavli Institute for Cosmological Physics, University of Chicago, Chicago, IL 60637, USA}
\affil{Department of Astronomy and Astrophysics, University of Chicago, Chicago, IL 60637, USA}

\author{C. ~E. ~Martínez-Vázquez}
\affil{Gemini Observatory, NSF’s NOIRLab, 670 N. A’ohoku Place, Hilo, HI 96720, USA}
\affil{ Cerro Tololo Inter–American Observatory, NSF’s NOIRLab, Casilla 603, La Serena, Chile}

\author{M.~Aguena}
\affil{Laborat\'orio Interinstitucional de e-Astronomia - LIneA, Rua Gal. Jos\'e Cristino 77, Rio de Janeiro, RJ - 20921-400, Brazil}

\author[0000-0002-7069-7857]{S.~S.~Allam}
\affil{Fermi National Accelerator Laboratory, P. O. Box 500, Batavia, IL 60510, USA}

\author{O.~Alves}
\affil{Department of Physics, University of Michigan, Ann Arbor, MI 48109, USA}

\author{E.~Bertin}
\affil{CNRS, UMR 7095, Institut d'Astrophysique de Paris, F-75014, Paris, France}
\affil{Sorbonne Universit\'es, UPMC Univ Paris 06, UMR 7095, Institut d'Astrophysique de Paris, F-75014, Paris, France}

\author[0000-0002-8458-5047]{D.~Brooks}
\affil{Department of Physics \& Astronomy, University College London, Gower Street, London, WC1E 6BT, UK}

\author[0000-0003-3044-5150]{A.~Carnero~Rosell}
\affil{Instituto de Astrofisica de Canarias, E-38205 La Laguna, Tenerife, Spain}
\affil{Laborat\'orio Interinstitucional de e-Astronomia - LIneA, Rua Gal. Jos\'e Cristino 77, Rio de Janeiro, RJ - 20921-400, Brazil}
\affil{Universidad de La Laguna, Dpto. Astrofísica, E-38206 La Laguna, Tenerife, Spain}

\author[0000-0002-3130-0204]{J.~Carretero}
\affil{Institut de F\'{\i}sica d'Altes Energies (IFAE), The Barcelona Institute of Science and Technology, Campus UAB, 08193 Bellaterra (Barcelona) Spain}

\author{L.~N.~da Costa}
\affil{Laborat\'orio Interinstitucional de e-Astronomia - LIneA, Rua Gal. Jos\'e Cristino 77, Rio de Janeiro, RJ - 20921-400, Brazil}

\author[0000-0001-8318-6813]{J.~De~Vicente}
\affil{Centro de Investigaciones Energ\'eticas, Medioambientales y Tecnol\'ogicas (CIEMAT), Madrid, Spain}

\author{P.~Doel}
\affil{Department of Physics \& Astronomy, University College London, Gower Street, London, WC1E 6BT, UK}

\author{I.~Ferrero}
\affil{Institute of Theoretical Astrophysics, University of Oslo. P.O. Box 1029 Blindern, NO-0315 Oslo, Norway}

\author{D.~Friedel}
\affil{Center for Astrophysical Surveys, National Center for Supercomputing Applications, 1205 West Clark St., Urbana, IL 61801, USA}

\author[0000-0003-4079-3263]{J.~Frieman}
\affil{Fermi National Accelerator Laboratory, P. O. Box 500, Batavia, IL 60510, USA}
\affil{Kavli Institute for Cosmological Physics, University of Chicago, Chicago, IL 60637, USA}

\author[0000-0002-9370-8360]{J.~Garc\'ia-Bellido}
\affil{Instituto de Fisica Teorica UAM/CSIC, Universidad Autonoma de Madrid, 28049 Madrid, Spain}

\author[0000-0002-3730-1750]{G.~Giannini}
\affil{Institut de F\'{\i}sica d'Altes Energies (IFAE), The Barcelona Institute of Science and Technology, Campus UAB, 08193 Bellaterra (Barcelona) Spain}

\author[0000-0003-3270-7644]{D.~Gruen}
\affil{University Observatory, Faculty of Physics, Ludwig-Maximilians-Universit\"at, Scheinerstr. 1, 81679 Munich, Germany}

\author{R.~A.~Gruendl}
\affil{Center for Astrophysical Surveys, National Center for Supercomputing Applications, 1205 West Clark St., Urbana, IL 61801, USA}
\affil{Department of Astronomy, University of Illinois at Urbana-Champaign, 1002 W. Green Street, Urbana, IL 61801, USA}

\author{S.~R.~Hinton}
\affil{School of Mathematics and Physics, University of Queensland,  Brisbane, QLD 4072, Australia}

\author{D.~L.~Hollowood}
\affil{Santa Cruz Institute for Particle Physics, Santa Cruz, CA 95064, USA}

\author[0000-0002-6550-2023]{K.~Honscheid}
\affil{Center for Cosmology and Astro-Particle Physics, The Ohio State University, Columbus, OH 43210, USA}
\affil{Department of Physics, The Ohio State University, Columbus, OH 43210, USA}

\author[0000-0003-0120-0808]{K.~Kuehn}
\affil{Australian Astronomical Optics, Macquarie University, North Ryde, NSW 2113, Australia}
\affil{Lowell Observatory, 1400 Mars Hill Rd, Flagstaff, AZ 86001, USA}

\author[0000-0001-9497-7266]{J. Mena-Fern{\'a}ndez}
\affil{Centro de Investigaciones Energ\'eticas, Medioambientales y Tecnol\'ogicas (CIEMAT), Madrid, Spain}

\author[0000-0002-1372-2534]{F.~Menanteau}
\affil{Center for Astrophysical Surveys, National Center for Supercomputing Applications, 1205 West Clark St., Urbana, IL 61801, USA}
\affil{Department of Astronomy, University of Illinois at Urbana-Champaign, 1002 W. Green Street, Urbana, IL 61801, USA}

\author[0000-0002-6610-4836]{R.~Miquel}
\affil{Instituci\'o Catalana de Recerca i Estudis Avan\c{c}ats, E-08010 Barcelona, Spain}
\affil{Institut de F\'{\i}sica d'Altes Energies (IFAE), The Barcelona Institute of Science and Technology, Campus UAB, 08193 Bellaterra (Barcelona) Spain}

\author[0000-0003-2120-1154]{R.~L.~C.~Ogando}
\affil{Observat\'orio Nacional, Rua Gal. Jos\'e Cristino 77, Rio de Janeiro, RJ - 20921-400, Brazil}

\author{M.~E.~S.~Pereira}
\affil{Hamburger Sternwarte, Universit\"{a}t Hamburg, Gojenbergsweg 112, 21029 Hamburg, Germany}

\author[0000-0001-9186-6042]{A.~Pieres}
\affil{Laborat\'orio Interinstitucional de e-Astronomia - LIneA, Rua Gal. Jos\'e Cristino 77, Rio de Janeiro, RJ - 20921-400, Brazil}
\affil{Observat\'orio Nacional, Rua Gal. Jos\'e Cristino 77, Rio de Janeiro, RJ - 20921-400, Brazil}

\author[0000-0002-2598-0514]{A.~A.~Plazas~Malag\'on}
\affil{Kavli Institute for Particle Astrophysics \& Cosmology, P. O. Box 2450, Stanford University, Stanford, CA 94305, USA}
\affil{SLAC National Accelerator Laboratory, Menlo Park, CA 94025, USA}

\author[0000-0002-9646-8198]{E.~Sanchez}
\affil{Centro de Investigaciones Energ\'eticas, Medioambientales y Tecnol\'ogicas (CIEMAT), Madrid, Spain}

\author{B.~Santiago}
\affil{Instituto de F\'\i sica, UFRGS, Caixa Postal 15051, Porto Alegre, RS - 91501-970, Brazil}
\affil{Laborat\'orio Interinstitucional de e-Astronomia - LIneA, Rua Gal. Jos\'e Cristino 77, Rio de Janeiro, RJ - 20921-400, Brazil}

\author[0000-0002-6261-4601]{J.~Allyn~Smith}
\affil{Austin Peay State University, Dept. Physics, Engineering and Astronomy, P.O. Box 4608 Clarksville, TN 37044, USA}

\author[0000-0002-3321-1432]{M.~Smith}
\affil{School of Physics and Astronomy, University of Southampton,  Southampton, SO17 1BJ, UK}

\author[0000-0002-7047-9358]{E.~Suchyta}
\affil{Computer Science and Mathematics Division, Oak Ridge National Laboratory, Oak Ridge, TN 37831}

\author[0000-0003-1704-0781]{G.~Tarle}
\affil{Department of Physics, University of Michigan, Ann Arbor, MI 48109, USA}

\author[0000-0001-7836-2261]{C.~To}
\affil{Center for Cosmology and Astro-Particle Physics, The Ohio State University, Columbus, OH 43210, USA}

\author{N.~Weaverdyck}
\affil{Department of Physics, University of Michigan, Ann Arbor, MI 48109, USA}
\affil{Lawrence Berkeley National Laboratory, 1 Cyclotron Road, Berkeley, CA 94720, USA}

\author{B.~Yanny}
\affil{Fermi National Accelerator Laboratory, P. O. Box 500, Batavia, IL 60510, USA}


\correspondingauthor{K.~B.~Webber}
\email{kbwebber@tamu.edu}

\begin{abstract}
We present a detailed chemical abundance analysis of the brightest star in the ultra-faint dwarf (UFD) galaxy candidate Cetus~II from high-resolution Magellan/MIKE spectra. For this star, DES J011740.53-173053, abundances or upper limits of 18 elements from Carbon to Europium are derived. Its chemical abundances generally follow those of other UFD galaxy stars, with a slight enhancement of the $\alpha$-elements (Mg, Si, and Ca) and low neutron-capture element (Sr, Ba, Eu) abundances supporting the classification of Cetus~II as a likely UFD. The star exhibits lower Sc, Ti, and V abundances than Milky Way (MW) halo stars with similar metallicity. This signature is consistent with yields from a supernova (SN) originating from a star with a mass of $\sim$11.2 M$_\odot$. In addition, the star has a Potassium abundance of $\mathrm{[K/Fe]} = 0.81$ which is somewhat higher than the K abundances of MW halo stars with similar metallicity, a signature which is also present in a number of UFD galaxies. A comparison including globular clusters (GC) and stellar stream stars suggests that high K is a specific characteristic for some UFD galaxy stars and can thus be used to help classify objects as UFD galaxies.
\end{abstract}


\section{Introduction \label{sec:intro}}
Stars maintain in their atmospheres a fingerprint of the chemical composition of their birth environment and thereby also contain information about the stellar generations that were present before them. They can therefore be used to study chemical enrichment, which is a useful probe to characterize the nature of the stellar environment. Variations in chemical abundances across different stellar associations can give insight into properties such as different timescales for chemical enrichment and variations in stellar initial mass functions.

In recent years, photometric surveys such as the Dark Energy Survey and Pan-STARRS \citep{DES, bechtol2015, koposov2015, drlicawagner2015, laevens2015} have detected a large number of stellar overdensities around the Milky Way (MW). In the process of characterizing these systems, chemical abundance analysis has proven to be a valuable tool to discern the physical nature of the associations and sometimes reveal unusual chemical abundance patterns. 
Although further observations of a few of these overdensities have determined that they are not genuine physical objects \citep{cantu2021}, nearly all of the overdensities have been found to be either star clusters or ultra-faint dwarf (UFD) galaxies \citep{laevens2014, luque, simon2019}.

 Star clusters can be sorted into different categories: stellar associations, open clusters, or GCs \citep{Trumpler, starcluster}. Each of these types of systems has distinguishing features in the abundances of their member stars that can be used to identify the type of cluster. Open clusters are homogeneous in the abundances of all elements, while GCs are not fully homogeneous \citep{starcluster}. GCs also are known to display specific anti-correlations between certain elements such as Mg-Al and Na-O \citep{carretta2010, Mucciarelli2018}. Hence, studying the chemical enrichment of star clusters can tell us useful information about the nature of these objects and aid in their classification.

 UFD galaxies are low mass, low luminosity, dark matter dominated galaxies \citep{simon2007}. The current stellar population of UFD galaxies is very old ($\sim$10~Gyr) and metal-poor ($\mathrm{[Fe/H]}< -1.4$, \citealt{simon2019}), with abundances  reflecting the elements created by the first generation of stars. Studying the abundances of stars in UFD galaxies, therefore, provides a window to study the nucleosynthetic processes of the early Universe. Furthermore, with the UFD galaxies being small, isolated systems, they reflect the chemical signatures of few nucleosynthetic events \citep{ji2016d, hansen2020, hansen2017, marshall2019}. 

Previous studies of stellar abundances in UFD galaxies have found generally similar abundance patterns from system to system \citep{frebel2015}, with most of the stars exhibiting a slight enhancement of the $\alpha$-elements (Mg, Si, Ca) \citep{simon2019, frebel2010, lai2011}, as seen for metal-poor stars in the MW halo \citep[e.g.,][]{mcwilliam1998, cayrel2004}, reflecting enrichment by core-collapse SN. In addition, and specific to UFD galaxies, most systems also display very low neutron-capture element abundances \citep[e.g.,][]{ji2019}. With the increase in the number of systems being discovered and analysed, more variation in the abundances has begun to appear. Examples of systems with unusual abundance patterns include Reticulum~II, where $\sim 72\%$ of stars analysed show enhancement in neutron-capture elements likely due to a neutron star merger event occurring early in the history of the galaxy \citep{ji2022,ji2016b,roederer2016}, and Grus~II, where all three stars analysed have a high $\mathrm{[Mg/Ca]}$ ratio compared to other UFD galaxies, pointing to a top heavy initial mass function governing the early star formation in this galaxy \citep{hansen2020}.

In this paper, we present a detailed chemical abundance analysis of the brightest star of the Cetus~II (Cet~II) UFD galaxy candidate, DES J011740.53-173053.1 (hereafter J0117). Cet~II was discovered as a stellar overdensity in the Dark Energy Survey, having a heliocentric distance of 30~kpc \citep{drlicawagner2015}. At the time of discovery, Cet~II was the faintest and smallest candidate UFD galaxy system detected, making it difficult to determine the velocity and metallicity dispersion. The brightest member was, therefore, a prime target for high-resolution follow-up aimed at abundance analysis. As described above, both UFD galaxies and star clusters have distinct chemical features that can be used to classify the system. Hence a detailed chemical analysis is the natural next step in the efforts to characterize the Cet~II system.

The outline of the paper is as follows. In Section \ref{sec:observations}, the observations are described, and in Section \ref{sec:stellar parameters and abundance analysis}, the stellar parameters and analysis are detailed. Section \ref{sec:results} presents our results that are further discussed in Section \ref{sec:discussion}. Section \ref{sec:summary} provides a summary.

\section{Observations \label{sec:observations}}

Based on medium-resolution spectroscopy of the Cet~II field obtained with the Magellan/IMACS spectrograph \citep{dressler2011} in 2016, we identified a set of likely Cet~II member stars centered at a velocity of $V_{\rm helio}=-82$~km~s$^{-1}$. Subsequently, \citet{Pace2019} and \citet{Pace2022} showed that these stars also share a common proper motion, confirming that they are associated with Cet~II. Surprisingly, given the low stellar mass of Cet~II, one of the member stars is located on the upper red giant branch at an apparent magnitude of $g=16.44$, $\sim3$~mag brighter than any other Cet~II stars. This bright star, J0117, was then an obvious target for high-resolution spectroscopy to investigate the chemical abundances in Cet~II. 

\begin{figure}
\centering
\includegraphics[scale=0.5]{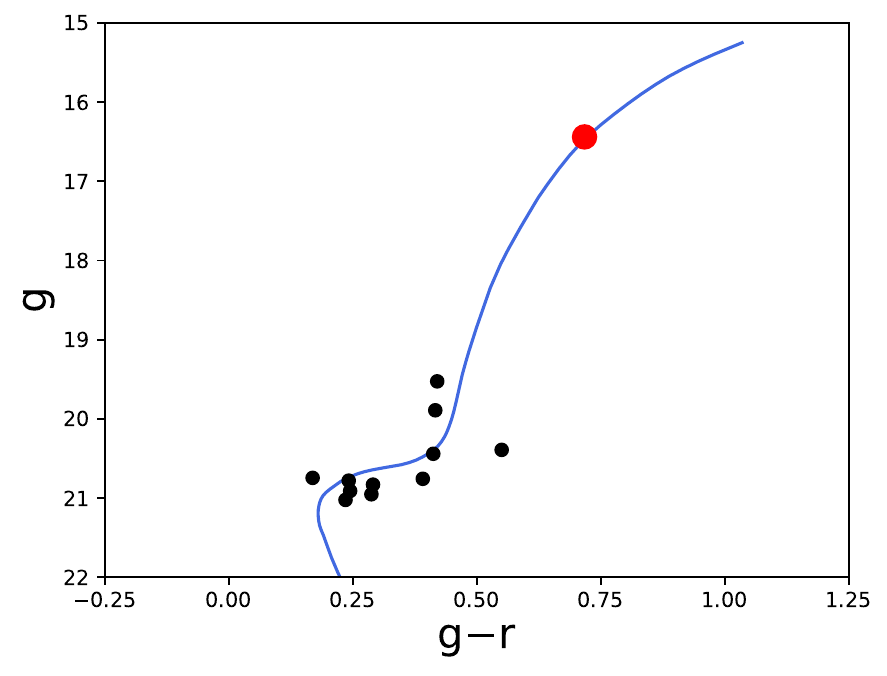}
\caption{\label{Fig:cmd} Color-magnitude diagram for Cetus~II. Black dots are stars with membership probability $p>0.01$ \citep{Pace2022} and red dot is J0117. Blue line shows a Dartmouth isochrone \citep{Dotter2008} with $\mathrm{[Fe/H]} = -2.3$ and age = 12.5~Gyr, transformed to the DES photometric system \citep{drlicawagner2018} and shifted to the distance of Cetus~II ($\sim$30 kpc, \citealt{drlicawagner2015}). }
\end{figure}

High-resolution spectral data was obtained for J0117 with the MIKE echelle spectrograph at Las Campanas Observatory in Chile \citep{bernstein2003} in August and November 2017. A color-magnitude diagram of Cet~II member stars \citep{Pace2022} is shown in Figure \ref{Fig:cmd}. J0117 is marked with a red dot and is notably brighter than the other currently known members. 

Right ascension, declination, and dereddened DES magnitudes for J0117 are listed in Table \ref{tab:obs}, along with Heliocentric Julian Date (HJD), exposure times, signal-to-noise ratio per pixel (S/N), and heliocentric radial velocities for the spectra. The spectra were obtained using a 0.7" slit with 2x2 pixel binning and cover a wavelength range of 3350-5000 \AA\  in the blue and 4900-9500 \AA\ in the red, with resolutions ($R$=$\lambda / \Delta\lambda$) of $R \sim$ 35,000 at blue wavelengths and $\sim28,000$ at red wavelengths, respectively. The data from each observing run were reduced using the CarPy MIKE pipeline \citep{kelson2000, kelson2003}, and the spectra from the two runs were subsequently co-added. The heliocentric radial velocity of the star was determined by cross-correlation with a spectrum of the bright, metal-poor red giant HD122563 ($V_{helio}$ = $-$26.17 km~s$^{-1}$; \citealt{gaiarv21}). Thirty-six echelle orders were used for the correlation in each spectrum, yielding mean radial velocities of  $-$81.38 km~s$^{-1}$ and $-$81.89 km~s$^{-1}$ for the two spectra. Since there is no appreciable variation in the radial velocities between the two spectra, this star is likely not a short-period binary.

\begin{deluxetable*}{lccccccccccc}

\caption{Observing Details and Stellar Data \label{tab:obs}}
\tablehead{ID & RA & Dec & $g_0$ & $r_0$ & $i_0$ & $z_0$ & HJD & Exp Time & S/N @ 4500\AA\ & $V_{helio}$ $\pm$ $\sigma$  \\ & & &(mag)&(mag)&(mag)&(mag) & & sec & & km~s$^{-1}$  &}
\startdata
 \multirow{2}{4em}{J0117} & \multirow{2}{4em}{01:17:40.9} & \multirow{2}{5em}{$-$17:30:54.5} & \multirow{2}{4em}{16.443} & \multirow{2}{4em} {15.716} & \multirow{2}{4em}{15.818}& \multirow{2}{4em}{16.014} &2457988.70127 & 2x1800  & 15 & $-$81.38 $\pm$ 0.52  \\
  &     &       &           &       &        &      &2458059.49903 & 5x1800 & 21 & $-$81.89 $\pm$ 0.51 
\enddata
\tablecomments{HJD values at beginning of exposure}
\end{deluxetable*}

\begin{deluxetable}{crrrrrrrrrrrrc}
\tablecaption{\label{tab:lines} Data for atomic lines used in analysis and individual line EW measurements.}
\tablehead{
\colhead{} & & & &  & \colhead{}\\
\colhead{Species} & \colhead{$\lambda$} & \colhead{$\chi$} & \colhead{$\log{gf}$} & \colhead{EW} &  \colhead{$\sigma_{\rm EW}$} & \colhead{$\log{\epsilon}$} &  \colhead{ref}\\
\colhead{} &\colhead{(\AA)} & \colhead{(eV)} & \colhead{} & \colhead{(m\AA)} & \colhead{(m\AA)} & \colhead{} &  & \colhead{}   }
\startdata
\ion{Na}{1} &  5889.95 &0.00&  0.11& 197.33&   2.83&  4.04 & 1\\
\ion{Na}{1} &  5895.92 &0.00& $-$0.19& 171.34&   2.89&  3.99 & 1\\
\ion{Mg}{1} &  4167.27 &4.35& $-$0.74& 84.82&  6.00&  5.80 & 1\\
\ion{Mg}{1} &  4702.99 &4.33&  $-$0.44&95.81&2.96&5.54 & 1\\
\ion{Mg}{1} &  5172.68 &2.71&  $-$0.36&248.53& 6.18&5.47 & 2\\

\enddata
\tablerefs{(1) \citet{kramida2018}; (2) \citet{pehlivan2017}; (3) \citet{yu2018}; (4) \citet{lawler1989}, using hfs from \citet{kurucz1995}; (5) \citet{lawler2013}; (6) = \citet{wood2013}; (7) \citet{pickering2001}, with corrections given in \citet{pickering2002}; (8) \citet{lawler2014} for $\log{gf}$ values and HFS; (9) \citet{wood2014a} for $\log{gf}$ values and HFS, when available; (10) \citet{sobeck2007}; (11) \citet{lawler2017}; (12) \citet{denhartog2011} for both $\log{gf}$ value and hfs; (13) \citet{obrian1991}; (14) \citet{denhartog2014}; (15) \citet{belmonte2017}; (16) \citet{ruffoni2014}; (17) \citet{denhartog2019}; (18) \citet{melendez2009}; (19) \citet{lawler2015} for $\log{gf}$ values and HFS; (20) \citet{wood2014b}; (21) \citet{roederer2012}; (22) \citet{kramida2018}, using HFS/IS from \citet{mcwilliam1998} when available.}
\tablecomments{The complete version of this Table is available online only. A short version is shown here to illustrate its form and content.}
\end{deluxetable}

\section{Stellar Parameters and Abundance Analysis \label{sec:stellar parameters and abundance analysis}}

Stellar parameters and abundances were derived from equivalent width (EW) measurements and spectral synthesis using the program Spectroscopy Made Hard(er) (\code{SMHR}\footnote{\url{https://github.com/andycasey/smhr}}), which runs the 2017 version of the radiative transfer code \code{MOOG}\footnote{\url{https://github.com/alexji/moog17scat}} \citep{sneden1973,sobeck2011} assuming local thermodynamical equilibrium (LTE). One dimensional (1D) $\alpha$-enhanced ([$\alpha$/Fe]= +0.4) ATLAS model atmospheres \citep{castelli2003} were used as input, with line lists generated from linemake which include hyperfine structure and isotopic shifts where applicable \footnote{\url{https://github.com/vmplacco/linemake}} \citep{placco2021} and Solar abundances were taken from \citet{asplund2009}. For the derivation of Ba, we used the $r$-process isotopic ratio from \citet{sneden2008}.

The stellar parameters effective temperature ($T_{\mathrm{eff}}$), surface gravity ($\log{g}$), metallicity ($\mathrm{[Fe/H]}$), and microturbulence ($\xi$) were determined spectroscopically using EW measurements of 95 \ion{Fe}{1} and 7 \ion{Fe}{2} lines (see Table \ref{tab:lines}). The EWs of the \ion{Fe}{1} and \ion{Fe}{2} lines were measured by fitting Gaussian profiles to the absorption features in the continuum-normalized spectra. Using these measurements, $T_{\mathrm{eff}}$ was derived from the excitation equilibrium of the \ion{Fe}{1} lines and then corrected for the offset between spectroscopic and photometric temperature scales using the method outlined in \citet{frebel2013}. Next, $\log{g}$ was determined from the ionization balance between the \ion{Fe}{1} and \ion{Fe}{2} lines, and $\xi$ was determined by removing any trend in line abundances with reduced equivalent width for the \ion{Fe}{1} lines. Final stellar parameters are listed in Table \ref{tab:params}. For comparison, a photometric temperature of 4592 $\pm$ 69~K for the star was also derived by converting the de-reddened DES $g$, $r$, $i$, and $z$ colors to $B$, $V$, $R$, and $I$ colors \citep[][R. Lupton 2005\footnote{http://www.sdss3.org/dr8/algorithms/sdssUBVRITransform.php}]{drlicawagner2018} and using the color-temperature relations from \citet{casagrande2010}. The photometric temperature is in good agreement with the corrected spectroscopic temperature. Following \citet{frebel2013}, we adopt a 150~K systematic uncertainty for $T_{\rm eff}$, corresponding to an uncertainty of 0.3 dex in $\log{g}$, 0.2 km~s$^{-1}$ in $\xi$, and 0.18 dex in $\mathrm{[Fe/H]}$. The statistical uncertainties on $T_{\rm eff}$, $\log{g}$ and $\xi$ were derived by varying each parameter to match the standard deviation of \ion{Fe}{1} lines as listed in Table \ref{tab:params}. Once the stellar parameters were determined, elemental abundances were derived from EW measurements or spectral synthesis. EWs were used for lines that are not blended, while spectral synthesis was used for blended lines and/or lines that are affected by isotopic and/or hyperfine splitting. Atomic data, wavelength, excitation potential, and oscillator strength for individual lines used for the abundance determination are listed in Table \ref{tab:lines}. The table also lists the measured EWs, uncertainties on these, and the corresponding abundances for the lines used. 
Final weighted mean abundances and associated uncertainties were determined following the method outlined in \citet{ji2020b}. This method uses a mean that is weighted by the S/N of the individual lines to calculate the final abundances and fully propagates statistical and systematic stellar parameter uncertainties for individual line measurements, including stellar parameter covariances, to determine the uncertainty.

\begin{deluxetable}{lcccc}
\caption{Stellar Model Atmosphere Parameters of J0117\label{tab:params}}
\tablehead{ & $T_{\rm eff}$ & $\log g$ & $\xi$ & $\mathrm{[Fe/H]}$ \\ & (K)&(cgs)&(km~s$^{-1}$)}
\startdata
 Value & 4727 & 1.40 & 1.89 & $-$2.09 \\
 Statistical uncertainties & 41 & 0.07 & 0.07 & 0.17 \\
 Systematic uncertainties & 150 & 0.3 & 0.2& 0.12
\enddata
\end{deluxetable}

\begin{deluxetable*}{lrrrrrrrrrrr}
\centerwidetable
\tablecolumns{12}
\tablecaption{\label{tab:abun}Weighted Average Abundance Summary for J0117 }
\tablehead{El. & N & $\log\epsilon$ & $\mathrm{[X/H]}$ & $\sigma_{\mathrm{[X/H]}}$ & $\mathrm{[X/Fe]}$ & $\sigma_{\mathrm{[X/Fe]}}$ & $\Delta_{T_{\rm eff}}$ & $\Delta_{\log g}$ & $\Delta_{\xi}$ & $\Delta_\mathrm{[Fe/H]}$ & $s_X$ 
 }
\startdata

\ion{CH} {0} &  4 & $+$5.85 &$-$2.61& 0.18& $-$0.31&  0.18 & 0.29&  $-$0.08&  0.01 &        0.10&  0.10\\	
\ion{Na}{1} &  2 & $+$4.02 &$-$2.12& 0.26& $+$0.17&  0.25 & 0.26&  $-$0.08&  $-$0.14&  $-$0.03&    0.00\\
\ion{Mg}{1} &  4 & $+$5.69 &$-$1.93& 0.17& $+$0.36&  0.18 & 0.15&  $-$0.07&  $-$0.07&  0.00&    0.18\\
\ion{Al}{1} &  2 & $+$3.79 &$-$2.63& 0.42& $-$0.33&  0.41 & 0.25&  $-$0.11&  $-$0.11&  $-$0.01&    0.46\\
\ion{Si}{1} &  3 & $+$5.86 &$-$1.71& 0.24& $+$0.59&  0.24 & 0.18&  $-$0.03&  $-$0.06&  $-$0.00&    0.28\\
\ion{K}{1}  &  2 & $+$3.50 &$-$1.49& 0.26& $+$0.81&  0.25 & 0.19&  $-$0.03&  $-$0.09&  $-$0.02&    0.09\\
\ion{Ca}{1} &  20 & $+$4.38 &$-$2.03& 0.10& $+$0.23&  0.11 & 0.15&  $-$0.03&  $-$0.06& $-$0.01&    0.11\\
\ion{Sc}{2} &  6 & $+$0.61 &$-$2.61& 0.19& $-$0.31&  0.19 & 0.04&   0.09&  $-$0.07&  0.03&    0.10\\
\ion{Ti}{1} &  14 & $+$2.64 &$-$2.52& 0.23& $-$0.23&  0.22 & 0.25&  $-$0.03&  $-$0.05&  $-$0.02&    0.23\\
\ion{Ti}{2} &  23 & $+$2.93 &$-$2.23& 0.17& $+$0.06&  0.18 & 0.06&   0.10&  $-$0.11&  0.04&    0.28\\
\ion{V}{1}  &  2 & $+$1.64 &$-$2.30& 0.19& $-$0.01&  0.18 & 0.17&  $-$0.02&  $-$0.01&  $-$0.04&    0.00\\
\ion{V}{2}  &  3 & $+$1.98 &$-$1.93& 0.17& $+$0.37&  0.18 & $-$0.02&  0.12&  $-$0.02&  0.03&    0.00\\
\ion{Cr}{1} &  9 & $+$3.13 &$-$2.59& 0.14& $-$0.30&  0.14 & 0.24&  $-$0.03&  $-$0.08&  $-$0.02&    0.08\\
\ion{Cr}{2} &  2 & $+$3.29 &$-$2.36& 0.12& $-$0.07&  0.13 & $-$0.03&  0.11&  $-$0.02&  0.02&    0.00\\
\ion{Mn}{1} &  5 & $+$2.97 &$-$2.48& 0.23& $-$0.18&  0.22 & 0.20&  $-$0.02&  $-$0.08&  $-$0.04&    0.10\\
\ion{Fe}{1} & 95 & $+$5.21 &$-$2.29& 0.08&\nodata &\nodata& 0.07&  $-$0.01&  0.04&  0.01&    0.24\\
\ion{Fe}{2} &  7 & $+$5.39 &$-$2.22& 0.15&\nodata &\nodata& $-$0.01&   0.12&  0.01&   0.03&    0.10\\
\ion{Co}{1} &  4 & $+$2.73 &$-$2.37& 0.32& $-$0.07&  0.31 & 0.25&  0.00&  $-$0.11&  $-$0.04&    0.00\\
\ion{Ni}{1} &  12 & $+$4.12 &$-$2.19& 0.13& $+$0.09&  0.13 & 0.17&  $-$0.01&  $-$0.04&  $-$0.01&    0.17\\
\ion{Zn}{1} &  1 & $+$2.27 &$-$2.29& 0.15& $+$0.01&  0.15 & 0.06&  0.07&  $-$0.02&  0.02&    0.00\\
\ion{Sr}{2} &  1 & $-$1.52 &$-$4.39& 0.43& $-$2.10&  0.43 & 0.07&   0.12&  $-$0.18&  0.01&    0.00\\
\ion{Ba}{2} &  1 & $-$2.34 &$-$4.52& 0.39& $-$2.23&  0.39 & 0.09&   0.10&  0.01&   0.09&    0.10\\
\ion{Eu}{2} &  1 & $<-$1.92 & $<-$2.44& \nodata& $<-$0.33& \nodata& \nodata& \nodata & \nodata & \nodata & \nodata\\

\enddata
\end{deluxetable*}

\section{Results \label{sec:results}}
\begin{figure*}
\centering
\includegraphics[width=\linewidth]{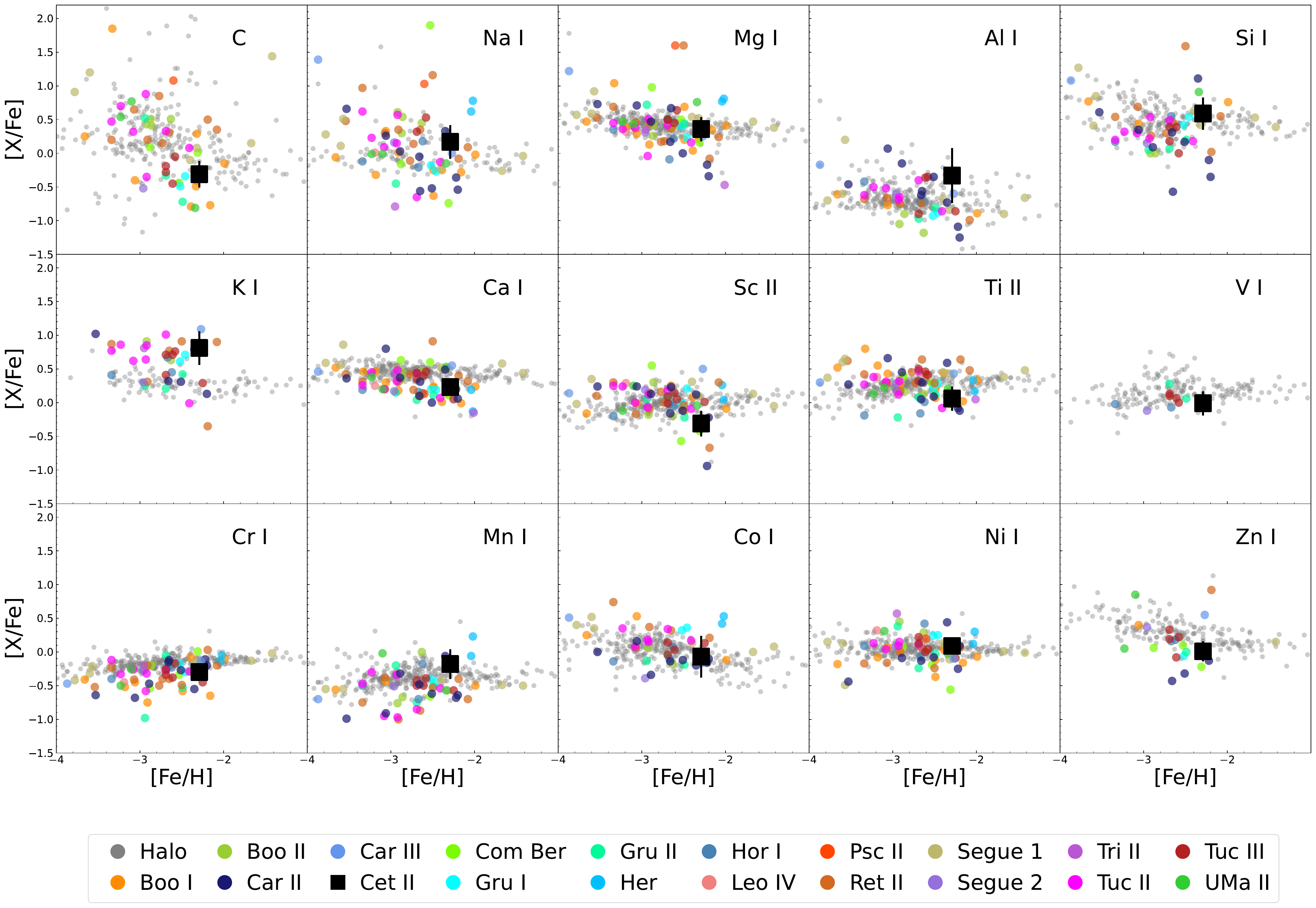}
\caption{\label{Fig:abun_tot} $\mathrm{[X/Fe]}$ derived abundances for J0117 (black square) compared to abundances of UFD galaxies (colored dots, see text for references \ref{sec:results}) and stars from the MW halo (gray dots \citealt{roederer2014}). J0117 follows the general abundance trends of what is seen in most UFD galaxy stars.  }
\end{figure*}

\begin{figure*}
\centering
\includegraphics[width=\linewidth]{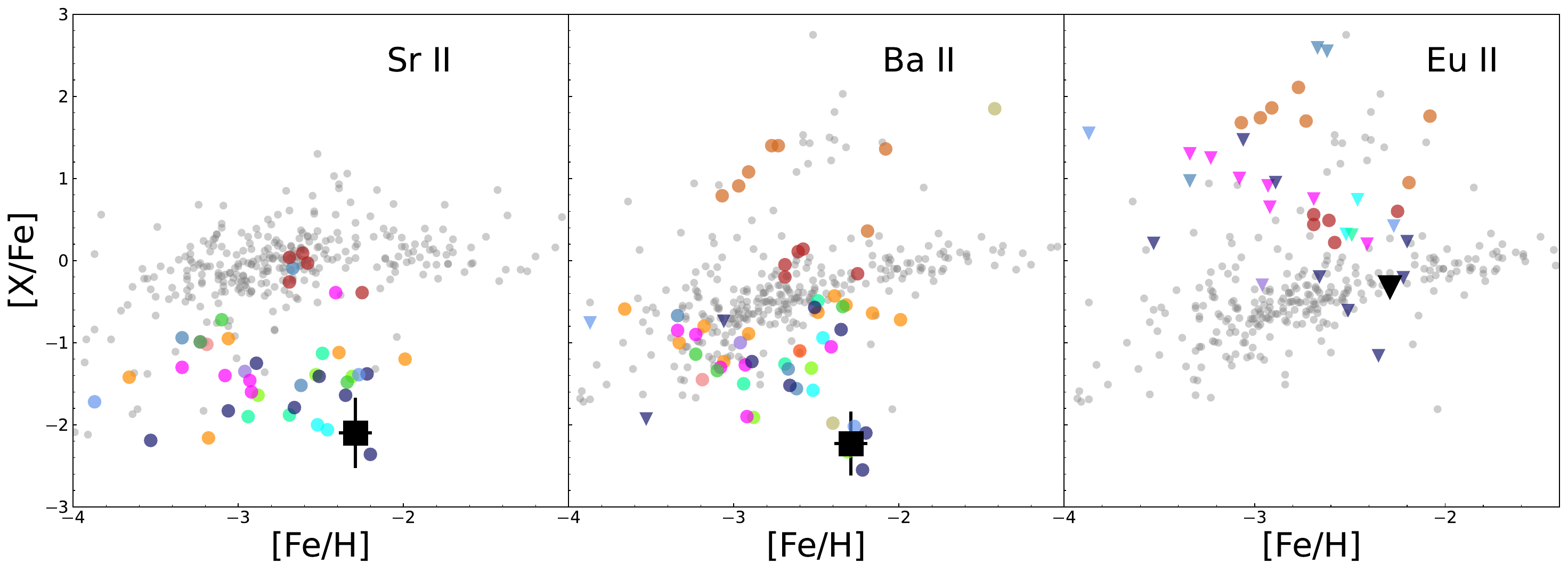}
\caption{\label{Fig:sr_ba_eu} $\mathrm{[X/Fe]}$ derived neutron-capture element abundances for J0117 (black square) compared to abundances of other UFD galaxies (colored dots, see text for references and Fig.~\ref{Fig:abun_tot} for legend) and stars from the MW halo (gray dots; \citealt{roederer2014}). Upper limits are designated with downward-pointing triangles. J0117 has low abundances of neutron-capture elements, which is characteristic of most UFD galaxies. }
\end{figure*}

Abundances or upper limits have been derived for 18 elements from C to Eu in J0117. The LTE abundances and upper limits, along with the systematic abundance uncertainty ($s_X$) and uncertainties arising from stellar parameter uncertainties ($\Delta_{{\mathrm T}_{eff}}$, $\Delta_{\log{g}}$, $\Delta_{\xi}$, and $\Delta_{[Fe/H]}$), are presented in Table \ref{tab:abun}, where N is the number of lines for the given species.  In Figure \ref{Fig:abun_tot} and \ref{Fig:sr_ba_eu}, we present a subset of elemental abundances for J0117 compared to abundances for stars in UFD galaxies (colored points) and metal-poor halo stars (grey points) \citep{roederer2014}. Data from the UFD galaxies are taken from: Bo\"{o}tes~I \citep{feltzing2009,frebel2016,gilmore2013,ishigaki2014,norris2010, waller2022}, Bo\"{o}tes~II \citep{ji2016a}, Carina~II \citep{ji2020a}, Carina~III \citep{ji2020a}, Coma Berenices \citep{frebel2010, waller2022}, Grus~I \citep{ji2019}, Grus~II \citep{hansen2020}, Hercules \citep{koch2008}, Horologium~I (Marshall et al. in prep.), Leo~IV \citep{simon2010}, Pisces~II \citep{spite2018}, Reticulum~II \citep{ji2016c,ji2019, hayes2023}, Segue~1 \citep{frebel2014,norris2010}, Segue~2 \citep{roederer2014a}, Triangulum~II \citep{ji2019}, Tucana~II \citep{ji2016d,chiti2018,chiti2023}, Tucana~III \citep{hansen2017,marshall2019}, and Ursa Major~II \citep{frebel2010}.

\subsection{Alpha elements}

We derive abundances for the $\alpha$-elements Mg, Si, and Ca from equivalent widths. We find a \ion{Mg}{1} abundance of $\mathrm{[Mg~I/Fe]} =  0.36 \pm{0.18}$ from four \ion{Mg}{1} absorption features, a \ion{Si}{1} abundance of $\mathrm{[Si~I/Fe]} = 0.59 \pm{0.24}$ using three \ion{Si}{1} lines, and a \ion{Ca}{1} abundance of $\mathrm{[Ca~I/Fe]}= 0.23 \pm{0.11}$ from 20 \ion{Ca}{1} lines. J0117 shows a general enhancement in $\alpha$-elements as a result of enrichment by core-collapse SN, following the trend seen in metal-poor MW halo and other UFD galaxy stars \citep{frebel2010,lai2011,simon2019}.

\subsection{Carbon and odd-Z elements}
Abundances for C, Na, Al, K, and Sc were derived from EW and spectral synthesis analysis. A $\mathrm{[C/Fe]}$ of $-$$0.31 \pm{0.18}$ was determined from the CH G-band in regions around 4310~\AA\ and 4230~\AA\ via spectral synthesis. The C and O abundances are coupled through the CO molecule, thus, a knowledge of the O abundance of the star is needed to derive the C abundance. As we could not derive an O abundance for the star, we assumed a standard $\alpha$-enhanced O abundance of $\mathrm{[O/Fe]} = 0.4$ for the spectral synthesis of CH, since UFDs are known to be $\alpha$-enhanced and J0117 is $\alpha$-enhanced. The C abundance of a star is also altered as the star evolves. Using the tool from \citet{placco2014}, a carbon correction of $\Delta\mathrm{[C/Fe]}$ = 0.57~dex was determined, resulting in an original C abundance of the star of $\mathrm{[C/Fe]} = 0.26$. 
A \ion{Na}{1} abundance of $\mathrm{[Na~I/Fe]} = 0.17 \pm{0.25}$ and an  \ion{Al}{1} abundance of $\mathrm{[Al~I/Fe]} = -0.33 \pm{0.41}$ were derived from EW analysis of the Na D resonance lines and two Al lines, respectively. EW analysis of the two K lines at 7664 and 7698~\AA\ was used to derive a \ion{K}{1} abundance of $\mathrm{[K~I/Fe]} = 0.81 \pm{0.25}$. From spectral synthesis of six \ion{Sc}{2} lines, we derive a \ion{Sc}{2} abundance of $\mathrm{[Sc~II/Fe]} = -0.31 \pm{0.19}$. 

As can be seen in Figure \ref{Fig:abun_tot}, the C and Na abundances for J0117 follow those of other UFD galaxies and metal-poor MW halo stars. The Al abundance for J0117 may be slightly higher than what is seen in other UFD galaxies, and metal-poor MW halo stars, but the uncertainty of $\pm 0.41$ is high. J0117 has an elevated K abundance compared to the average K abundances found in MW halo stars. In addition, J0117 exhibits a lower Sc abundance than what is found in most UFD galaxies and metal-poor MW halo stars. 

\subsection{Iron peak elements}
EW analysis was used to derive abundances for the iron peak elements Ti, Cr, Ni, and Zn and spectral synthesis for the elements V, Mn, and Co. We identify 14 \ion{Ti}{1} and 23 \ion{Ti}{2} lines in the spectrum and find abundances of $\mathrm{[Ti~I/Fe]} = -0.23 \pm{0.22}$ and $\mathrm{[Ti~II/Fe]} = 0.06 \pm{0.18}$. \ion{V}{1} and \ion{V}{2} abundances were determined to be $\mathrm{[V~I/Fe]} = -0.01 \pm{0.18}$ and $\mathrm{[V~II/Fe]} = 0.37 \pm{0.18}$ from two and three lines respectively and \ion{Cr}{1} and \ion{Cr}{2} to be $\mathrm{[Cr~I/Fe]} = -0.30 \pm{0.14}$ and $\mathrm{[Cr~II/Fe]} = -0.07 \pm{0.13}$ from nine and two lines respectively.
There is a slight discrepancy between the abundances we derive from the neutral and ionized lines of the elements mentioned above, which may be due to the 1D LTE nature of the analysis. However, for most of the elements, the neutral and ionized abundances agree within uncertainties. Finally, we derive \ion{Ni}{1} and \ion{Zn}{1} abundances of $\mathrm{[Ni~I/Fe]} = 0.09 \pm{0.13}$ and $\mathrm{[Zn~I/Fe]} = 0.01 \pm{0.15}$ using 12 Ni lines and one Zn line. Comparing these abundances to those of other UFD galaxy and MW halo stars (see Figure \ref{Fig:abun_tot}) reveals the Cet~II star to have relatively low Ti and V abundances, while the remainder of the iron peak element abundances for this star are similar to other UFD galaxy and metal-poor MW halo stars.

\subsection{Neutron-capture elements}
\begin{figure}
\centering
\includegraphics[scale=0.5]{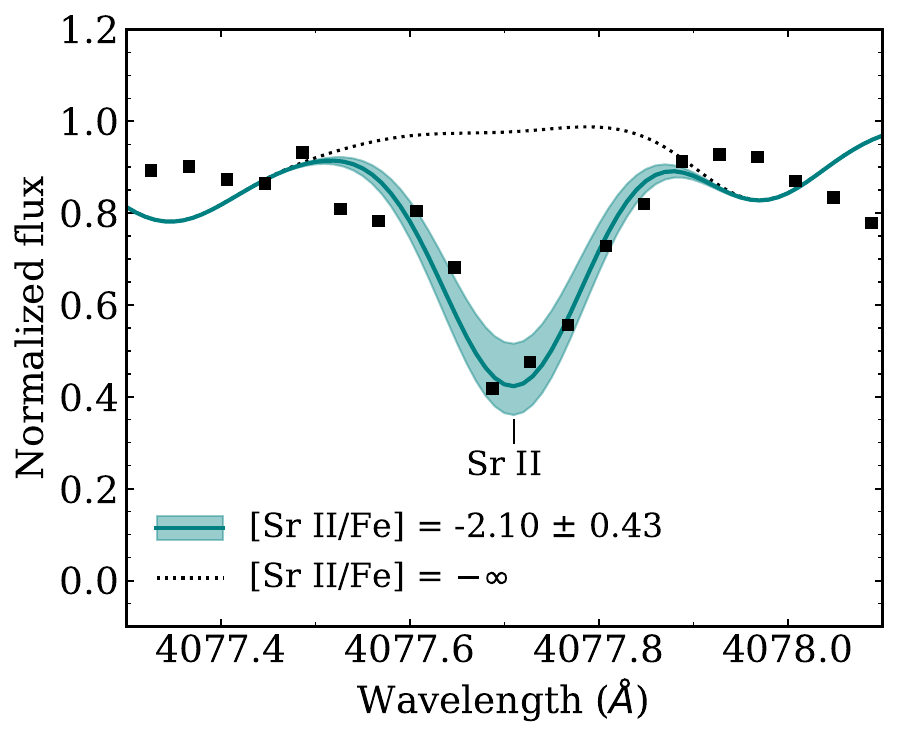}\\
\includegraphics[scale=0.5]{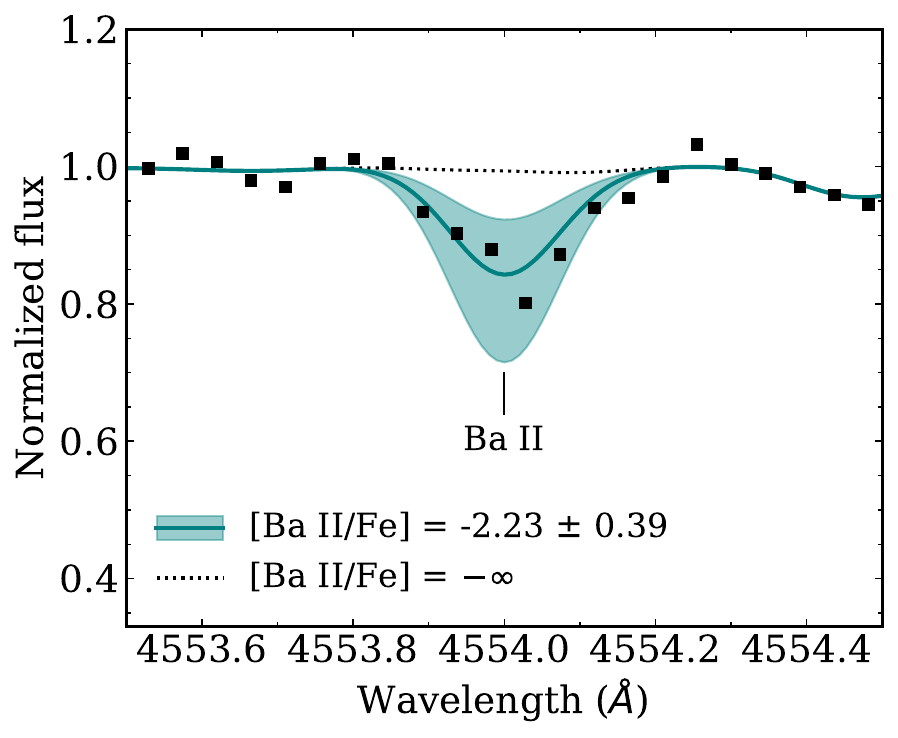}
\caption{Spectral synthesis of the 4077 \AA\ Sr~II line (top) and 4554 \AA\ Ba~II line (bottom). Black squares are the observed spectrum of J0117. The dotted line shows a synthesis not including Sr or Ba, and the blue line shows a synthesis with the $\mathrm{[Sr/Fe]} \pm \sigma_{\mathrm{[Sr/Fe]}}$ or $\mathrm{[Ba/Fe]} \pm \sigma_{\mathrm{[Ba/Fe]}}$ value derived for J0117.\label{Fig:sr_synthesis}  }
\end{figure}

Absorption features from the two neutron-capture elements Sr and Ba were identified in the spectrum, and using spectral synthesis, we derive abundances for these elements of $\mathrm{[Sr~II/Fe]} = -2.10 \pm{0.43}$ and $\mathrm{[Ba~II/Fe]} = -2.23 \pm{0.39}$. The synthesis of the Sr 4077~\AA\ and Ba 4554~\AA\ lines are shown in Figure \ref{Fig:sr_synthesis}. The Sr abundance derived for J0117 is low and similar to what is found in most other UFD galaxies, and the Ba abundance is lower than the majority of other UFD galaxy stars, see Figure \ref{Fig:sr_ba_eu}. No Eu could be detected in the spectrum, so we derive a 3$\sigma$ upper limit of $\mathrm{[Eu~II/Fe]}<-0.33$ from the 4129~\AA\, line, plotted as a downwards pointing triangle in Figure \ref{Fig:sr_ba_eu}.

\section{Discussion \label{sec:discussion}}

\subsection{Nature of Cetus~II}
The nature of the Cet~II overdensity is still debated in the literature. The DES discovery paper \citep{drlicawagner2015} and a more recent paper from the DES based on analysis of more extensive photometry \citep{drlicawagner2020} both classify it as a probable UFD galaxy, while \citet{conn2018}, who presented GMOS-S photometry of the system, suggested it is more likely a part of the Sagittarius (Sgr) tidal stream. However, the proper motion of the Cet~II system, $(\mu_{\alpha \cos{\delta}}, \mu_\delta)=(+2.8\pm{0.06}, 0.5 \pm{0.06})~{\rm mas~yr^{-1}}$ \citep{Pace2019, Pace2022} does not agree with the proper motion of the Sgr stream $(\mu_{\alpha \cos{\delta}}, \mu_\delta)=(-1, -3)~{\rm mas~yr^{-1}}$ \citep{Vasiliev2021} and therefore rules out an association with the Sgr stream. As well as this, the velocity measured for J0117 does not agree with the velocity of the Sgr stream \citep{Vasiliev2021} which further confirms that they are not associated. This will be explored more in forthcoming work (Simon et al. in prep.).

The size of Cet~II ($r_{1/2} \sim$17pc) \citep{drlicawagner2015} lies in an ambiguous region in the size-luminosity plane where the star cluster and UFD populations overlap \citep{willmanstrader2012}. As discussed in \cite{willmanstrader2012}, the most effective classification for these systems is a velocity dispersion followed by a metallicity dispersion. These have not yet been measured for Cet~II. 

In this paper, we have derived the abundances of one star, J0117, in Cet~II, and the general pattern of this star is similar to other UFD galaxy stars (see Figure \ref{Fig:abun_tot}) and hence compatible with a UFD galaxy classification of Cet~II. This is particularly supported by the very low abundance of the neutron-capture elements Sr and Ba, which have been found to be characteristic of UFD galaxies and can be used to classify a system \citep{ji2019}, in particular, this abundance feature distinguishes UFD galaxy stars from GC stars. Furthermore, we suggest that a high K abundance is another potential UFD abundance signature, which will be explored further in Section \ref{sec:High K abundance}. No abundance analysis exists for Sgr stream stars at metallicities similar to J0117; however, there have been abundance analyses for stars in the Sgr dwarf spheroidal galaxy (dSph) at similar metallicities to J0117. The Sgr dSph stars show abundance signatures that we do not find in J0117, including higher neutron capture element abundances ($-0.2< \mathrm{[Sr/Fe]} <0.6$,$-0.8 <\mathrm{[Ba/Fe]} <0.8$, $-0.17< \mathrm{[Eu/Fe]} <0.81$ \citealt{hansen2018,reichert2019}). Thus, the chemical abundances of Cet~II do not support an association with the Sgr dSph.  For the remainder of the discussion, we will adopt the classification of Cet~II as a UFD galaxy.

\subsection{Abundance pattern of J0117}
With its low metallicity, slight enhancement of the $\alpha$-elements, and low abundances of neutron-capture elements, J0117 generally follows the same abundance trends seen in the majority of UFD galaxy stars. However, the derived abundances for some elements stand out and warrant further discussion.

\subsubsection{Low Sc, Ti, and V abundances}
\begin{figure*}
\centering
\includegraphics[scale=0.5]{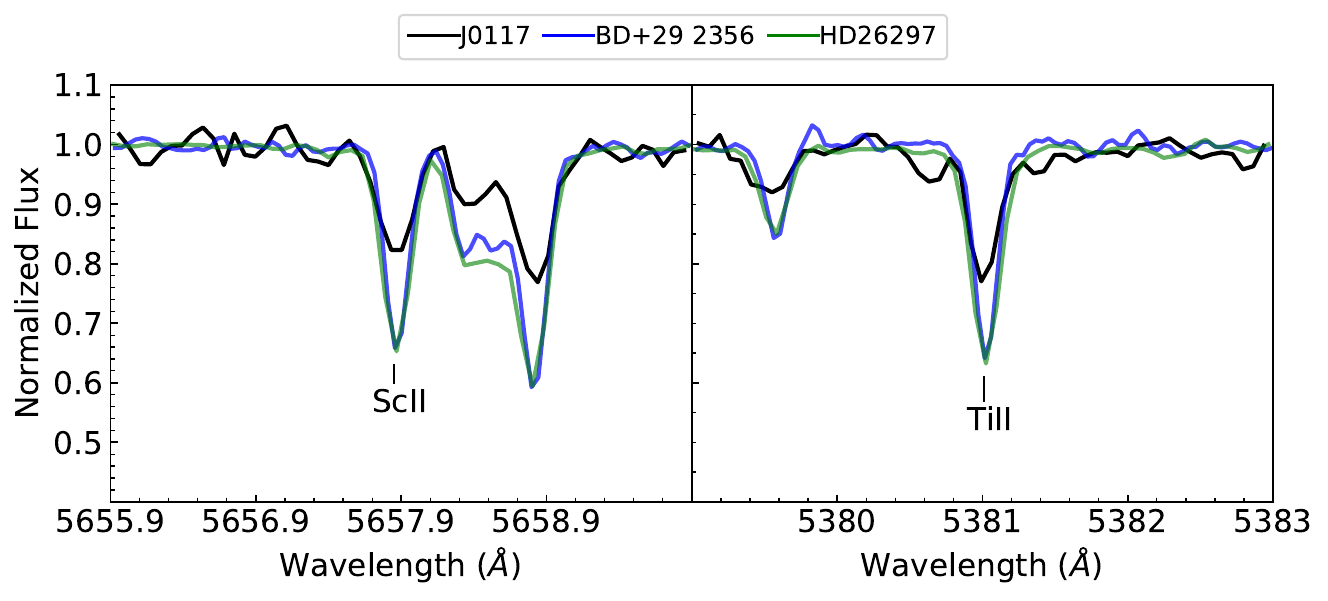}
\caption{\label{Fig:comparison} { Comparison of \ion{Sc}{2} and \ion{Ti}{2} absorption lines in J0117 to the MW halos stars BD+29 2356, and HD~26297 from \citet{roederer2014}. The absorption features for J0117 are weaker than what is seen for BD+29 2356 and HD~26297. Model atmosphere parameters and Sc and Ti abundances for each star are shown in Table \ref{tab:comp}.}}
\end{figure*}

One specific feature of J0117 is that the $\mathrm{[Sc/Fe]}$, $\mathrm{[Ti/Fe]}$, and $\mathrm{[V/Fe]}$ ratios are $\sim 0.3, 0.2,$ and $0.1$ dex lower, respectively, in this star than the average of what is seen in most other UFD galaxy stars and metal-poor MW halo stars (see Figure \ref{Fig:abun_tot}). We plot the 5657 \AA\ \ion{Sc}{2} and 5381 \AA\ \ion{Ti}{2} lines in the spectrum of J0117 (black) compared to spectra of stars HD 26297 (green) and BD+29 2356 (blue) (I. Roederer private comm.), in Figure \ref{Fig:comparison}. HD 26297 and BD+29 2356 were chosen because they have similar model atmosphere parameters to J0117 but higher Sc and Ti abundances. However, it should be noted that there is a difference in $\mathrm{[Fe/H]}$ which amplifies the intrinsic difference in $\mathrm{[Sc/Fe]}$ and $\mathrm{[Ti/Fe]}$. The parameters for each star and abundances for each line are listed in Table \ref{tab:comp}. It is clear from Figure \ref{Fig:comparison} that the absorption features of Sc and Ti are weaker in the spectrum of J0117, thus supporting the lower abundances derived for these elements in this star. 

\begin{deluxetable*}{lcccccc}
\scriptsize
\caption{Model Atmosphere Parameters and Sc and Ti Abundances for Comparison Stars and J0117.\label{tab:comp} } 

\tablehead{ID & $T_{\rm eff}$ & $\log g$ & $\xi$ & Model $\mathrm{[Fe/H]}$ &  $\log\epsilon$(Sc II) &  $\log\epsilon$(Ti II) \\ &  (K)&(cgs)&(km~s$^{-1}$)}
\startdata
 HD~26297 & 4400 & 1.10 & 1.75 & -1.72 & 1.57 & 3.67 \\
 BD +29 2356 & 4710 & 1.75 & 1.50 & -1.62 &1.65 & 3.61  \\
 J0117 & 4727 & 1.40 & 1.89 & -2.09 & 0.63 & 3.00 
\enddata
\tablecomments{Abundances for BD +29 2356 and HD~26297 are taken from \citet{roederer2014}.}
\end{deluxetable*}

The Sc, Ti, and V abundances for metal-poor stars have been investigated by \citet{sneden2016, cowan2020}, and \citet{vanadium}, who all found that the abundances of these three elements are positively correlated in metal-poor stars ($\mathrm{[Fe/H]} < -2$), indicating that these three elements have linked nucleosynthetic origins. While the main focus of \citet{cowan2020} is the analysis of UV spectra taken with the Hubble Space Telescope of three metal-poor stars, they also investigated the correlation among Sc, Ti, and V abundances derived in several large spectroscopic studies \citep[e.g.,][]{cayrel2004,cohen2004,cohen2008,barklem2005,lai2008,yong2013,roederer2014}, and found that the abundances reported in these studies also show correlations between the three elements.

\begin{figure}
\centering
\includegraphics[scale=0.45]{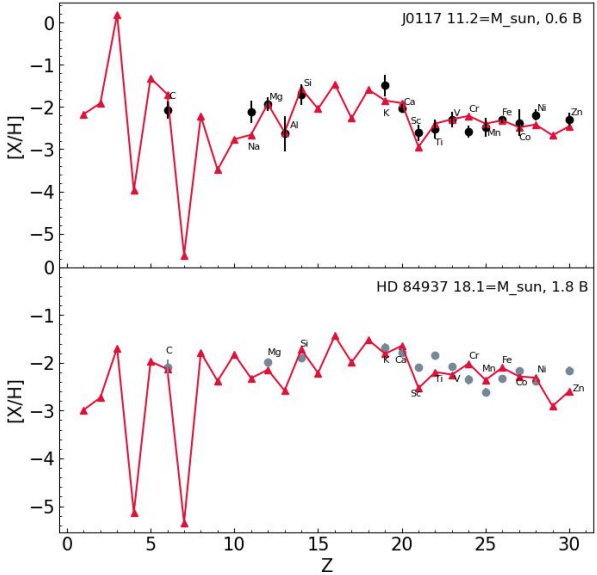}
\caption{\label{Fig:starfit} Best-fit SN yield model \citep{Heger2010} for derived abundances in J0117 (top) and HD~84937 (bottom). Black points are the derived abundances for J0117, gray points are the derived abundances for HD~84937 \citep{sneden2016, spite2017}, and the red line is the best-fit SN yield model for each star.}
\end{figure}

Ti and V are produced via explosive Si- and O- burning in core-collapse SN \citep{woosley1995,sneden2016}, with models producing similar V/Ti ratios for both the Si and O SN ejecta components \citep[e.g.,][]{pignatari2016}. \citet{sneden2016} suggested that the correlation of the Ti and V abundances in metal-poor stars points to co-production of these elements and that their abundances can be used to explore properties of the progenitor and its explosion, such as mass and explosion energy. The case of Sc is a bit more complicated. Although it can be produced in the same explosive Si- and O-burning layers \citep{woosley1995}, it is more efficiently produced via neutrino feedback or during $\alpha$-rich freeze-out conditions \citep{woosley1995,frolich2006}. This fact led \citet{sneden2016} to argue that, although a correlation is seen, the Sc/Ti and Sc/V ratios are less useful diagnostics than V/Ti ratios of the properties of the progenitor star. Generally, theoretical SN models often underproduce all three elements compared to the abundances derived from metal-poor stars \citep[][and references therein]{kobayashi2020}. However, it was suggested by both \citet{cowan2020} and \citet{vanadium} that more energetic SN, like hypernovae, might contribute to the abundances of these elements. 

Contrary to the star analyzed in \citet{sneden2016} and the three stars presented in \citet{cowan2020}, which all display higher than normal Sc, Ti, and V abundances, J0117 exhibits lower than average abundances for these elements. Since the chemistry of these stars is a direct fingerprint of the nucleosynthesis process of the first massive (Population III) stars to form in the galaxies, they can be used to constrain properties such as mass range, rotation, and explosion energies of the Pop III stars. To investigate the signature of J0117, we used the \code{STARFIT} tool\footnote{\url{https://starfit.org/}} to match SN yields from \citet{Heger2010} to the abundances of J0117. The \code{STARFIT} code calculates a $\chi^2$ statistic using the derived abundances and upper limits to determine the best-fit SN yields. For J0117, we obtained models with $\chi^2_{red}<2$, which returned models in a narrow range of progenitor masses, M$_{\odot}$=10.6-13.6. The best fit ($\chi^2_{red}$= 1.12) was achieved with a progenitor mass of 11.2~M$_{\odot}$ and a modest explosion energy of 0.6~B \footnote{B=1 Bethe = $10^{51}$ erg}. Figure \ref{Fig:starfit} shows the abundances for J0117 and best-fit model yields. This model provides a good fit to most elements, with only a few discrepancies. It should be noted that it is unlikely that this star was polluted by a singular Pop III SN, but this tool gives an idea of the most dominant source of metals. To demonstrate the range of masses that can come from these abundances, we also derived the best-fit yields for the star from \citet{sneden2016}, HD~84937. HD~84937 is used for comparison because it has high Sc, Ti, and V abundances compared to the average halo stars and thus represents the high end of the abundance distribution, with J0117 representing the low end. Abundances for the light elements were taken from \citet{spite2017}, who adopted similar stellar parameters for HD~84937 as \citet{sneden2016}. For HD~84937, the best fit was achieved with a progenitor mass of 18.1~M$_{\odot}$ and explosion energy of 1.8 B. This demonstrates how the abundances for these elements can be used to gain information about the progenitor masses and explosion energies as well as how these parameters can vary with different abundances. Furthermore, the progenitor mass found for the Cet~II star is also somewhat lower than what has been found for stars in other UFD galaxies \citep{hansen2020}, pointing to a wide range of progenitor masses existing in UFD galaxies.

Another possibility to explain the low Sc and V abundances that we see is pollution from Type Ia SN. As seen in \citet{bravo2019}, the production of Sc and V in Type Ia SN is metallicity dependent and decreases with decreasing metallicity. Hence, it is possible that the low Sc and V abundances are a result of increased Fe injection from Type Ia SN with low Sc and V production. However, the $\alpha$-element abundance ratios in J0117 are slightly enhanced and compatible with core-collapse SN production and don't suggest extra Fe injection from Type Ia SN.

\subsubsection{High K abundance \label{sec:High K abundance}}
\begin{figure*}
\centering
\includegraphics[scale=0.5]{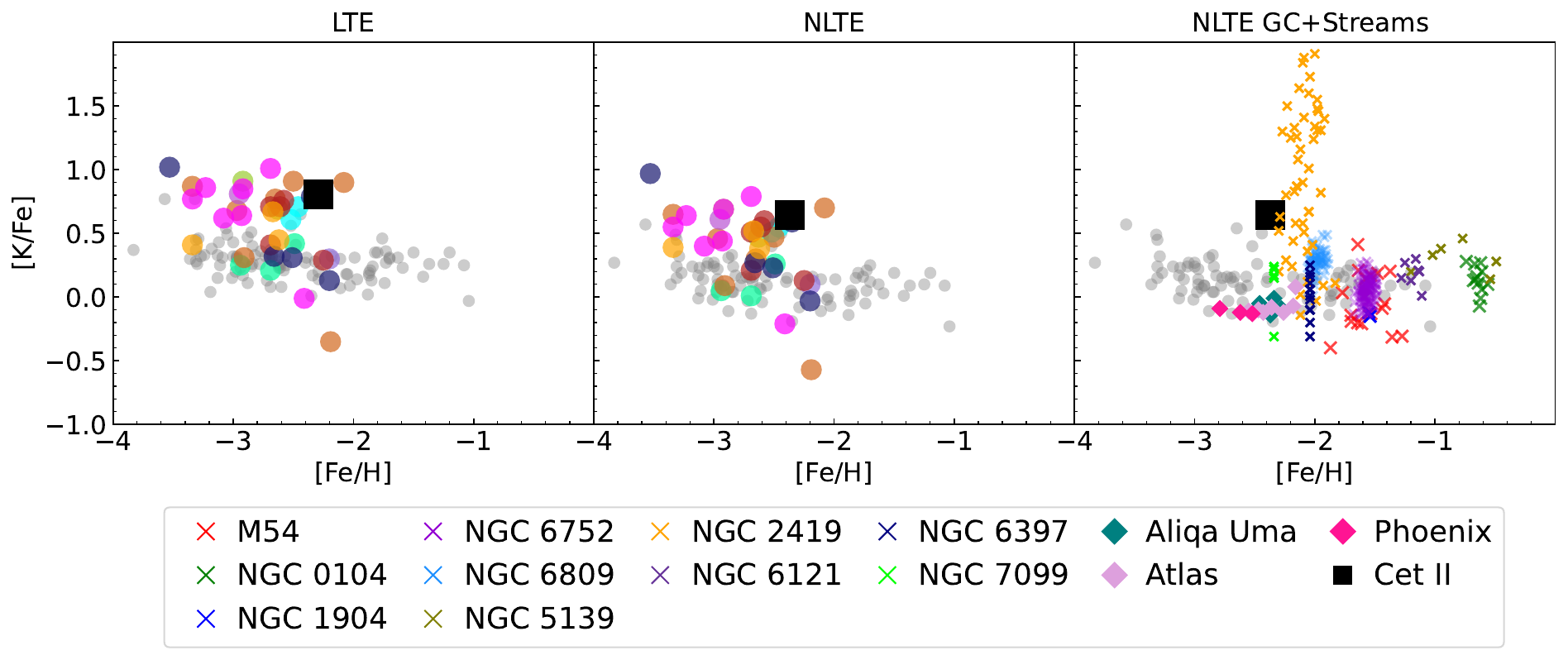}
\caption{\label{Fig:potassium}Left: LTE $\mathrm{[K/Fe]}$ abundances for J0117 (black square), UFDs (colored dots, see Fig.\ref{Fig:abun_tot} for legend), and MW halo (gray dots) stars. Middle: NLTE corrected $\mathrm{[K/Fe]}$ abundances for J0117, UFDs, and MW halo stars. Right: NLTE corrected $\mathrm{[K/Fe]}$ abundances for J0117, MW halo stars, GC stars (colored x), and stars from stellar streams with GC progenitor (colored diamonds). There is an offset in $\mathrm{[K/Fe]}$ abundance between some UFD galaxies and the average value for MW halo stars. Also, the GC and stellar stream $\mathrm{[K/Fe]}$ abundances mostly overlap with the MW halo stars and are lower than what is seen in UFD galaxy stars.}
\end{figure*}

Another peculiar feature of J0117 is the somewhat high $\mathrm{[K/Fe]}$ abundance of 0.81 derived for this star. But, as can be seen in Figure \ref{Fig:abun_tot}, J0117 is not a complete outlier. High K has been found in several other UFD galaxy stars. In fact, in Figure \ref{Fig:abun_tot}, it can be seen that stars in UFD galaxies generally have higher $\mathrm{[K/Fe]}$ ratios than MW halo stars at similar metallicity. However, K is known to suffer from non-local thermodynamic equilibrium (NLTE) effects \citep{mueller1975}. We, therefore, corrected the K abundance derived for J0117 and the K abundances of the UFD galaxy literature sample following the study of \citet{reggiani2019} and got a $\mathrm{[K/Fe]}$ abundance of $0.64$ for J0117. The NLTE-corrected K abundances for J0117 and other UFD galaxy stars, along with NLTE-corrected K abundances of halo stars \citep{roederer2014}, are shown in the middle panel of Figure \ref{Fig:potassium}, while the LTE abundances are shown in the panel on the left. It can be seen that the offset between a subset of the UFD galaxies and the halo stars remains after the correction is applied. \cite{ivanova2000} also looked at the NLTE corrections for K and provided corrections slightly larger than \cite{reggiani2019}. However, even with these corrections, the offset between the MW halo stars and a subset of the UFD galaxies remains.

Apart from the NLTE effects, the K lines also have hyperfine structure. However, this is very weak and, therefore, generally not included in the abundance analysis. But for completeness, we derived K abundances from the two lines used above via spectral synthesis, including hyperfine structure. As expected, the effect on the abundances was minimal (<0.1). Hence, for compatibility with the literature data, the discussion below is based on our K abundance derived from the EW analysis.

Although the abundances of J0117 are compatible with Cet~II being a UFD galaxy, we searched the literature for K abundances in GCs to explore if this abundance signature is unique to stars in UFD galaxies or can be found in other types of small stellar systems. The results of this exercise are plotted in the right panel of Figure \ref{Fig:potassium}, where the GC NLTE-corrected K abundances for stars in ten GCs (M54 \citep{Carretta2022}, NGC~2419 \citep{Mucciarelli2012},  NGC~104, NGC~6809 \citep{Mucciarelli2017}, NGC~6752 \citep{carretta2007}, NGC~1904, NGC~5139, NGC~6121, NGC~6397, NGC~7099 \citep{Carretta2013}) are marked with crosses. As can be seen, for most GCs, the K abundances overlap with the halo stars, suggesting that the high K abundances seen in UFD galaxy stars may be a unique abundance signature of these systems. The exception is NGC~2419, in which stars with a wide range of K abundances are found, including K-rich ($\mathrm{[K/Fe]} > 1$) stars. However, the K-rich stars in this cluster are all Mg-poor ($\mathrm{[Mg/Fe]}$ < 0) \citep{cohen2012,Mucciarelli2012}, thus making them distinguishable from the UFD galaxy stars, which usually exhibit a small enhancement in Mg (J0117, for example, has $\mathrm{[Mg/Fe]} = 0.36$). Although we find K abundances in a number of GCs, most have metallicities higher than the UFD galaxy stars with K abundances, hampering a direct comparison. However, in recent years a number of stellar streams have been detected that are thought to be the remnants of more metal-poor GCs accreted by the MW, such as the Phoenix, Aliqa Uma, and ATLAS streams \citep{li2019,wan2020,casey2021,li2022}. \citet{ji2020b} derived K abundances for stars in the three GC streams listed above. These abundances are plotted as diamonds in Figure \ref{Fig:potassium}. Similar to the present-day GCs, the stream stars also have K abundances overlapping with the MW halo stars. Hence, high K abundances seem to be a characteristic abundance signature for at least some UFD galaxies.

The nucleosynthetic origin of the high K abundances seen in UFD galaxy stars is yet unknown. K is created through hydrostatic oxygen shell burning and explosive oxygen burning with yields depending on the progenitor mass \citep{woosley1995}. Recently, models have shown that K production is increased when rotation is introduced \citep{prantzos}. Hence, the high K abundances in UFD galaxy stars could suggest that some of these stars were enriched by massive rotating stars. The models also show low Sc production in massive rotating stars \citep{prantzos}, which is in agreement with the Sc in J0117 ($\mathrm{[Sc/Fe]=-0.31}$). However, it is not a good fit for the other UFD stars that do not have low Sc. 

Since the high K abundances are mainly seen in the UFD galaxy stars, it is possible that this signature is tied to the slow chemical evolution of these systems. Although unlikely given the overall abundance signature of J0117, the K could also come from a nucleosynthesis source with a time-delayed contribution, like Type Ia SN. One candidate could be Ca-rich transients \citep{kasliwal2012}, the lightcurves of which \citet{polin2021} recently found to match models of sub-Chandrasekhar Type Ia SNe with low mass progenitors. However, more modeling is needed to establish if these events can contribute to the K abundances of UFD galaxy stars.

\section{Summary \label{sec:summary}}
We have performed a detailed chemical abundance analysis of J0117, the brightest star of the Cet~II UFD galaxy candidate. Our analysis shows that this star is a metal-poor, $\alpha$-enhanced ($\alpha$-elements $\gtrapprox 0.4$) star with low abundances of the neutron-capture elements ($\mathrm{[Sr/Fe]}=-2.10$ and $\mathrm{[Ba/Fe]}=-2.23$), following the trends seen for chemical analysis of other UFD galaxy stars. Thus, although there is still debate on the classification of this system, the results of this chemical analysis suggest Cet~II is a UFD galaxy. Further observations and chemical analysis of more stars in this system will help to fully determine the nature of Cet~II. The analysis revealed that the star exhibits slightly lower $\mathrm{[Sc/Fe]}$, $\mathrm{[Ti/Fe]}$, and $\mathrm{[V/Fe]}$ abundances compared to other UFD galaxy and MW halo stars at similar metallicities. It has been suggested that the abundances for $\mathrm{[Sc/Fe]}$, $\mathrm{[Ti/Fe]}$, and $\mathrm{[V/Fe]}$ can be used as a diagnostic for progenitor mass; we thus compared the abundances of J0117 with the \citet{Heger2010} SNe yields and determined a best-fit with a progenitor mass of 11.2M$_\odot$, somewhat lower than the progenitor masses found for stars in other UFD galaxies \citep{hansen2020}. Finally, we derive a K abundance of $\mathrm{[K/Fe]} = 0.81$ for J0117, which, even after it has been corrected for NLTE effects, is somewhat higher than the K abundances derived for MW halo stars at similar metallicities. We note that a number of UFD galaxies have high K abundances compared to the MW halo stars, and by including K abundances for stars in GCs and streams in our comparison, we find that this is a unique signature of some UFD galaxy stars.

\acknowledgements
The authors would like to thank Dr. Maria Drout for collecting the spectra of J0117 from November 2017 and Dr. Andrew McWilliam for atomic K line data and insightful comments on K abundances in stars.

This research made extensive use of the SIMBAD database operated at CDS, Straasburg, France \citep{wenger2000}, \href{https://arxiv.org/}{arXiv.org}, and NASA's Astrophysics Data System for bibliographic information.

Funding for the DES Projects has been provided by the U.S. Department of Energy, the U.S. National Science Foundation, the Ministry of Science and Education of Spain, 
the Science and Technology Facilities Council of the United Kingdom, the Higher Education Funding Council for England, the National Center for Supercomputing 
Applications at the University of Illinois at Urbana-Champaign, the Kavli Institute of Cosmological Physics at the University of Chicago, 
the Center for Cosmology and Astro-Particle Physics at the Ohio State University,
the Mitchell Institute for Fundamental Physics and Astronomy at Texas A\&M University, Financiadora de Estudos e Projetos, 
Funda{\c c}{\~a}o Carlos Chagas Filho de Amparo {\`a} Pesquisa do Estado do Rio de Janeiro, Conselho Nacional de Desenvolvimento Cient{\'i}fico e Tecnol{\'o}gico and 
the Minist{\'e}rio da Ci{\^e}ncia, Tecnologia e Inova{\c c}{\~a}o, the Deutsche Forschungsgemeinschaft and the Collaborating Institutions in the Dark Energy Survey. 

The Collaborating Institutions are Argonne National Laboratory, the University of California at Santa Cruz, the University of Cambridge, Centro de Investigaciones Energ{\'e}ticas, 
Medioambientales y Tecnol{\'o}gicas-Madrid, the University of Chicago, University College London, the DES-Brazil Consortium, the University of Edinburgh, 
the Eidgen{\"o}ssische Technische Hochschule (ETH) Z{\"u}rich, 
Fermi National Accelerator Laboratory, the University of Illinois at Urbana-Champaign, the Institut de Ci{\`e}ncies de l'Espai (IEEC/CSIC), 
the Institut de F{\'i}sica d'Altes Energies, Lawrence Berkeley National Laboratory, the Ludwig-Maximilians Universit{\"a}t M{\"u}nchen and the associated Excellence Cluster Universe, 
the University of Michigan, NSF's NOIRLab, the University of Nottingham, The Ohio State University, the University of Pennsylvania, the University of Portsmouth, 
SLAC National Accelerator Laboratory, Stanford University, the University of Sussex, Texas A\&M University, and the OzDES Membership Consortium.

Based in part on observations at Cerro Tololo Inter-American Observatory at NSF's NOIRLab (NOIRLab Prop. ID 2012B-0001; PI: J. Frieman), which is managed by the Association of Universities for Research in Astronomy (AURA) under a cooperative agreement with the National Science Foundation.

The DES data management system is supported by the National Science Foundation under Grant Numbers AST-1138766 and AST-1536171.
The DES participants from Spanish institutions are partially supported by MICINN under grants ESP2017-89838, PGC2018-094773, PGC2018-102021, SEV-2016-0588, SEV-2016-0597, and MDM-2015-0509, some of which include ERDF funds from the European Union. IFAE is partially funded by the CERCA program of the Generalitat de Catalunya.
Research leading to these results has received funding from the European Research
Council under the European Union's Seventh Framework Program (FP7/2007-2013) including ERC grant agreements 240672, 291329, and 306478.
We  acknowledge support from the Brazilian Instituto Nacional de Ci\^encia
e Tecnologia (INCT) do e-Universo (CNPq grant 465376/2014-2).

This manuscript has been authored by Fermi Research Alliance, LLC under Contract No. DE-AC02-07CH11359 with the U.S. Department of Energy, Office of Science, Office of High Energy Physics.
T.T.H acknowledges support from the Swedish Research Council (VR 2021-05556)

\facility{Magellan:Clay}
\software{MOOG \citep{sneden1973,sobeck2011}, IRAF \citep{tody1986,tody1993}, ATLAS9 \citep{castelli2003}, linemake \citep{placco2021}, NumPy \citep{numpy}, Matplotlib \citep{matplotlib}, AstroPy \citep{Astropy:13,Astropy:18}, CarPy \citep{kelson2003}, 
SMHR \citep{casey2014}}

\end{document}